\documentclass[final,5p,times]{elsarticle}
\usepackage{amsmath,amssymb,amsfonts}
\usepackage{graphicx}
\usepackage{slashed}
\usepackage{a4wide}
\usepackage{fancyhdr}
\usepackage{subfigure}

\allowdisplaybreaks[1]

\newcommand{\beq}{\begin{equation}}
\newcommand{\eeq}{\end{equation}}
\newcommand{\beqa}{\begin{eqnarray}}
\newcommand{\eeqa}{\end{eqnarray}}

\begin{document}

\renewcommand{\theequation}{\arabic{equation}}

\begin{frontmatter}

\title{Neutral pion photoproduction off $^3$H and $^3$He
in chiral perturbation theory}

\author[Bonn]{Mark~Lenkewitz}
\author[Bochum]{Evgeny~Epelbaum}
\author[Bonn]{H.-W.~Hammer}
\author[Bonn,Juelich]{Ulf-G.~Mei{\ss}ner}

\address[Bonn]{Helmholtz--Institut f\"ur Strahlen- und Kernphysik (Theorie) 
   and Bethe Center for Theoretical Physics, Universit\"at Bonn,\\ D-53115 Bonn,
   Germany}
\address[Bochum]{Institut f\"ur Theoretische Physik II, Ruhr-Universit\"at Bochum, D-44870, Germany}
\address[Juelich]{Institut f\"ur Kernphysik (Theorie), Institute for Advanced Simulation, 
   and J\"ulich Center for Hadron Physics,\\ Forschungszentrum J\"ulich, D-52425  J\"ulich, Germany}

\begin{abstract}
We calculate electromagnetic neutral pion production off three-nucleon bound
states ($^3$H, $^3$He) at threshold to leading one-loop order
in the framework of chiral nuclear effective
field theory. In addition, we analyze the dependence of the nuclear S-wave amplitude
$E_{0+}$ on the elementary neutron amplitude $E_{0+}^{\pi^0n}$ which in the case
of $^3$He provides a stringent test of the prediction based on chiral perturbation 
theory. Uncertainties from higher order corrections are estimated. 
\end{abstract}


\end{frontmatter}

\section{Introduction}

Threshold neutral pion photo- and electroproduction off the nucleon is one
of the finest reactions to test the chiral QCD dynamics, see 
\cite{Bernard:2007zu}
for a recent review. While elementary proton targets are accessible 
directly in experiment,
pion production off neutrons requires the use of nuclear targets
like the deuteron or three-nucleon bound states like $^3$H (triton) or $^3$He.
For a recent review on revealing the neutron structure from electron or photon
scattering off light nuclei, see \cite{Phillips:2009af}.
Of particular interest is to test the counterintuitive chiral perturbation
theory prediction (CHPT) that the elementary neutron S-wave multipole 
$E_{0+}^{\pi^0
  n}$ is larger in magnitude than the corresponding one of the proton,  
 $E_{0+}^{\pi^0 p}$ \cite{Bernard:1994gm,Bernard:2001gz}. This prediction
was already successfully tested in neutral pion photo- \cite{Beane:1997iv} 
and electroproduction off the deuteron \cite{Krebs:2004ir}. 
However, given the
scarcity and precision of the corresponding data, it is mandatory to study
also pion production off three-nucleon bound states, that can be calculated
nowadays to high precision based on chiral nuclear effective field theory
(EFT), that extends CHPT to  nuclear physics (for a recent review, see
\cite{Epelbaum:2008ga}). $^3$He appears to be a particularly promising target
to extract the information about the neutron amplitude. Its wave function is
strongly dominated by the principal ``s-state'' component which suggests that 
the spin of $^3$He is largely driven by the one of the neutron.  
Consequently, in this  letter we calculate
thres\-hold pion photo- and electroproduction  based on chiral 3N wave
functions at next-to-leading order in the chiral expansion. Experimentally,
neutral pion photoproduction off light nuclei has so far only been studied
at Saclay~\cite{Argan:1980zz,Argan:1987dm}  and at Saskatoon 
\cite{Bergstrom:1998zz,Barnett:2008zz}.

\begin{figure}[ht]
        \begin{center}
        \includegraphics[width=0.7\linewidth]{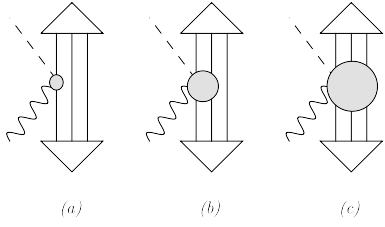}
        \end{center}
        \vspace{-0.2cm}
        \caption{Different topologies contributing to pion
        production off the three-nucleon bound state (triangle).
        (a), (b) and (c) represent the single-, two- and three-nucleon
        contributions, respectively. Solid, dashed and wiggly lines
        denote nucleons, pions and photons, in order. Topology (c) 
        does not contribute to the order considered here (NLO).}
        \label{fig:123N}
        \end{figure}

In general, one has three different topologies for pion production off a 
three-nucleon bound state as shown in Fig.~\ref{fig:123N}.
While the single-nucleon contribution (a) features the
elementary neutron and proton production amplitudes, the
nuclear corrections are given by two-body (b) and three-body 
(c) terms.  Based on the power counting developed in \cite{Beane:1995cb},
at next-to-leading order (NLO), only the to\-po\-lo\-gies (a) and (b)
contribute.  Here, we will specifically consider threshold photo- and
electroproduction parameterized in terms of the electric $E_{0+}$
and longitudinal $L_{0+}$ S-wave multipoles. In particular, we
will study the sensitivity of the three-body S-wave multipoles
to the elementary  $E_{0+}^{\pi^0 n}$ multipole, taking the proton
amplitude  $E_{0+}^{\pi^0 p}$ from CHPT (as this value is consistent 
with the data \cite{Schmidt:2001vg} and a recent study based on a 
chiral unitary approach \cite{Gasparyan:2010xz}).

\section{Anatomy of the Calculation}

To analyze the process under consideration, we calculate
the  the nuclear matrix  element of the given
transition operator $\hat{O}$ as:
\beq\label{eq:OME}
\langle M_J^\prime|\hat{O}|M_J\rangle_{\psi}
:=\langle \psi M_J^\prime \vec{P}^\prime_{3N}\vec{q}_\pi|\hat{O}|\psi 
M_J \vec{P}_{3N}\vec{k}_\gamma \rangle\, ,
\eeq
where $\psi$ refers to the three-nucleon wave function and
$\vec{k}_\gamma$, $\vec{q}_\pi$, $\vec{P}_{3N}$ and $\vec{P}^\prime_{3N}$
denote the
momentum of the exchanged (virtual) photon, produced pion and the initial
and final momentum of the  3N nucleus, respectively. 
The $3N$ bound state has total nuclear angular momentum $J=1/2$ with magnetic
quantum numbers $M_J$ for the initial and $M_J^\prime$ for the final nuclear 
state. $J$ can be decomposed in total spin $S=1/2,3/2$ and total orbital
angular momentum $L=0,1,2$. The total isospin is a mixture of two components, 
$T=1/2$ and $3/2$. While the $T=1/2$ component is large, the small $T=3/2$
component emerges due to isospin breaking and is neglected in  
our calculation. 
Here, we consider neutral pion production by real or virtual photons
of a spin-1/2 particle - either the nucleon or the $^3$H and $^3$He nuclei.
At threshold, the corresponding transition matrix takes the form
\beq
  \label{eq:multidef}
  \mathcal{M}_\lambda =
  2i\, E_{0+} \, (\vec{\epsilon}_{\lambda, \text{T}}\cdot\vec{S} )
  + 2i \, L_{0+}\, (\vec{\epsilon}_{\lambda, \text{L}}\cdot\vec{S}) \,,
\eeq
with $\vec{\epsilon}_{\lambda, \text{T}} = \vec{\epsilon}_\lambda
-(\vec{\epsilon}_\lambda\cdot\hat{k}_\gamma)\hat{k}_\gamma$ and $\vec{\epsilon}_{\lambda,
  \text{L}} = (\vec{\epsilon}_\lambda\cdot\hat{k}_\gamma)\hat{k}_\gamma$ the transverse and
longitudinal photon polarization vectors. The transverse and 
longitudinal  S-wave multipoles are denoted by $ E_{0+}$ and $L_{0+}$,
respective\-ly. Note that  $L_{0+}$  contributes only for virtual photons.

As explained before, the matrix element Eq.~(\ref{eq:OME}) receives
contributions from one- and two-nucleon operators at the order we are working. 
Consider first the single 
nucleon contribution, given in terms of the 1-body transition operator
$\hat{O}^{\rm 1N}$. After some algebra, one finds
\begin{align}\label{eq:O1fin}
&\langle M_J^\prime|\hat{O}^{\rm 1N}|M_J\rangle_{\psi} \nonumber\\
& =
i \vec{\epsilon}_{\lambda,\text{T}}\cdot\vec{S}_{M_J^\prime
  M_J}\Big(  E_{0+}^{\pi^0p} F_T^{S+V} + E_{0+}^{\pi^0n}  F_T^{S-V}\Big)\nonumber\\
&+ i \vec{\epsilon}_{\lambda,\text{L}}\cdot\vec{S}_{M_J^\prime
  M_J}\Big( L_{0+}^{\pi^0p}
F_L^{S+V}+ L_{0+}^{\pi^0n} F_L^{S-V}\Big)~.
\end{align}
where $F_{T/L}^{S\pm V}\equiv F_{T/L}^S\pm F_{T/L}^V$ and 
$F_{T/L}^{S,V}$ denote the corresponding form factors of the
3N bound state,
\begin{align}
\label{eq:F1N}
F_{T/L}^S \, \vec{\epsilon}_{\lambda,\text{T/L}}\cdot\vec{S}_{M_J^\prime
  M_J}
&= \frac{3}{2} \,  \langle M_J^\prime
| \vec{\epsilon}_{\lambda,\text{T/L}}\cdot\vec{\sigma}_{1}
|
M_J\rangle_{\psi}~, \\
F_{T/L}^V  \, \vec{\epsilon}_{\lambda,\text{T/L}}\cdot\vec{S}_{M_J^\prime
  M_J}
&= \frac{3}{2} \, \langle M_J^\prime
|\vec{\epsilon}_{\lambda,\text{T/L}}\cdot\vec{\sigma}_{1}\tau_{1}^z|
M_J\rangle_{\psi}~, \nonumber
\end{align}
which parametrize the overall normalization of the
response of the
composite system to the excitation by photons in spin-isospin space.
In the above equation, $\vec \sigma_{i}$ ($\vec \tau_i$) denote the spin
(isospin) Pauli matrices corresponding to the nucleon $i$. Furthermore, $z$
refers to the isospin quantization axis. 

Using the 3N wave functions from chiral nuclear EFT at the appropriate order,
the pertinent matrix elements in Eq.~(\ref{eq:F1N}) can be evaluated. 
Here, we use  chiral 3N wave functions obtained from the N$^2$LO
interaction in the Weinberg power counting
\cite{Epelbaum:2003gr,Epelbaum:2003xx}.\footnote{The consistency of the
  Weinberg counting for short-range operators and
  the non-perturbative renormalization of chiral EFT 
  are currently under discussion, see the 
  review~\cite{Epelbaum:2008ga} for more details.
  A real alternative to the Weinberg approach for practical
  calculations, however, is not available.} 
In order to estimate the 
error from higher order corrections, we use wave functions
for five different combinations of the 
cutoff $\tilde{\Lambda}$ in the spectral function regularization 
of the two-pion exchan\-ge and the cutoff $\Lambda$ used to regularize the 
Lipp\-mann-Schwinger equation for the two-body T-ma\-trix.
The wave functions are taken from Ref.~\cite{Liebig:2010ki,Noggaprivate}
and the corresponding cutoff combinations in units of MeV are 
$(\tilde{\Lambda},\Lambda)$ = (450,500), (600,500),
(550,600), (450,700), (600,700).
All five sets describe the binding energies of the 
$^3$He and $^3$H nuclei equally well.

The one-body contributions to the 3N multipoles are given by
\begin{align}\label{eq:mult1N}
E_{0+}^{\rm 1N}
& = \frac{K_{\rm 1N}}{2} \left( E_{0+}^{\pi^0 p} \, F_T^{S+V} + E_{0+}^{\pi^0 n}
\, F_T^{S-V}\right)~,\nonumber\\
L_{0+}^{\rm 1N}
& = \frac{K_{\rm 1N}}{2} \left( L_{0+}^{\pi^0p} \, F_L^{S+V} + L_{0+}^{\pi^0n}
\, F_L^{S-V} \right)~.
\end{align}
Here, $K_{1N}$ is the kinematical factor to account for the change in phase
space from the 1N to the 3N system, 
\begin{equation}\label{eq:kine1N}
K_{\rm 1N}
 = \frac{m_N+M_\pi}{m_{\rm 3N}+M_\pi}\frac{m_{\rm 3N}}{m_N}
\approx 1.092\,,
\end{equation}
with $m_N$ being the nucleon mass and 
$m_{\rm 3N}$ the mass of the three-nucleon bound state.

We evaluate the matrix elements for the one-body contribution
in Eq.~(\ref{eq:F1N}) numerically
with Monte Carlo integration using the VEGAS algorithm~\cite{Vegas}.
The results for the form factors $F_{T/L}^{S\pm V}$ are given
in Table~\ref{tab:FTL}.
\renewcommand{\arraystretch}{1.3}
\begin{table}[t]
\begin{center}
\begin{tabular}{|c||c|c|}
\hline 
nucleus & $^3$He & $^3$H \\
\hline
$F_{T}^{S+V}$ &  $0.017(13)(3)$ & $1.493(25)(3)$ \\ 
$F_{T}^{S-V}$ &  $1.480(26)(3)$ & $0.012(13)(3)$ \\
$F_{L}^{S+V}$ &  $-0.079(14)(8)$ & $1.487(27)(8)$ \\
$F_{L}^{S-V}$ &  $1.479(26)(8)$ & $-0.083(14)(8)$ \\
\hline
\end{tabular}
 \caption{Numerical results for the form factors $F_{T/L}^{S\pm V}$.
   The first error is an estimate of the theory error
   from higher orders in chiral EFT while the second error
   is the statistical error from the Monte Carlo integration.}
  \label{tab:FTL}
\end{center}
\end{table}
The first error represents the theoretical uncertainty estimated from the 
cutoff variation
in the wave functions. We take the central value defined by the five
different cutoff sets as our prediction and estimate the theory error
from higher-order corrections
from the spread of the calculated values. Strictly speaking, this procedure
gives a lower bound on the error, but in practice it generates a 
reasonable estimate. 

We stress that we follow the nuclear EFT formulation of Lepage, 
in which the whole effective potential  is iterated to all orders when 
solving the Schr\"odinger equation for the nuclear states.
As discussed in Ref.~\cite{Lepage:1997cs}, the cutoff should be kept
of the order of the breakdown scale or below
in order to avoid unatural scaling 
of the coefficients of higher order terms. Indeed, using larger cutoffs
can lead to a violation of certain low-energy theorems as demonstrated in 
Ref.~\cite{Epelbaum:2009sd} for an exactly solvable model.

The error related to the expansion of the
production operator is difficult to estimate given that the convergence in the
expansion for the single nucleon S-wave multipoles is known to be slow, see 
Ref.~\cite{Bernard:1994gm} for an extended discussion.
We therefore give here only a rough estimate of this uncertainty. The
extractions of the proton S-wave photoproduction amplitude based
on CHPT using various approximations \cite{FernandezRamirez:2009jb} 
lead to an uncertainty $\Delta E_{0+}^{\pi^0 p} \approx \pm 0.05\times
10^{-3}/M_{\pi^+}$, which is about 5\%. Similarly, we estimate the
uncertainty  of the neutron S-wave threshold amplitude to be the same.
Consequently, our estimate of the error on the single nucleon amplitude
is 5\%.

The statistical error from the evaluation of the
integrals is typically one order of magnitude 
smaller than the estimated theory error and can be neglected.

\begin{figure}[t]
        \begin{center}
        \includegraphics[width=0.8\linewidth]{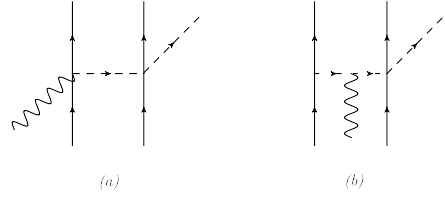}
        \end{center}
        \vspace{-0.2cm}
        \caption{Leading two-nucleon contributions to the 
         nuclear pion production matrix element at threshold.
        Solid, dashed and wiggly lines
        denote nucleons, pions and photons, in order.}
        \label{fig:abterm}
        \end{figure}

We now switch to the two-nucleon contribution. 
In Coulomb gauge, only the two Feynman diagrams shown in 
Fig.~\ref{fig:abterm} contribute at threshold to the order
we are working~\cite{Beane:1997iv}.
%
%
Their contribution to the multipoles can be written as
\beqa\label{eq:mult2N}
E_{0+}^{\rm 2N} &=& K_{\rm 2N} \, (F_T^{(a)} -F_T^{(b)})~,\nonumber\\
L_{0+}^{\rm 2N} &=& K_{\rm 2N} \, (F_L^{(a)} -F_L^{(b)})~,
\eeqa
with the prefactor
\begin{eqnarray}
K_{\rm 2N} &=&  \frac{M_\pi e g_A m_{\rm 3N}}{16\pi(m_{\rm 3N}+M_\pi)
(2\pi)^3 F_\pi^3}\nonumber\\
&\approx& 0.135\mbox{ fm }\times 10^{-3}/M_{\pi^+}~.
\end{eqnarray}
The numerical value for $K_{\rm 2N}$ 
was obtained using $g_A=1.26$ for the axial 
coupling constant, $F_\pi=93$ MeV for the pion decay constant, and the 
neutral pion mass $M_\pi=135$ MeV.
The transverse and longitudinal form factors $F_{T/L}^{(a)}$
and $F_{T/L}^{(b)}$ corresponding to diagrams (a) and (b), respectively, 
are
\begin{align}
&F_{T/L}^{(a)} \, \vec{\epsilon}_{\lambda,\text{T/L}}\cdot\vec{S}_{M_J^\prime
  M_J}  \nonumber\\
&=  \frac{3}{2}\, 
\langle M_J'|\frac{\vec{\epsilon}_{\lambda,\text{T/L}}\cdot (\vec \sigma_1
  + \vec \sigma_2) 
(\vec{\tau}_1\cdot\vec{\tau}_2-
\tau_1^z\tau_2^z)}{\left(\vec{p}_{12}-\vec{p}_{12}^{\;\prime}+
\vec{k}_\gamma/2\right)^2}|
M_J\rangle_\psi~, \nonumber\\
\label{eq:valF2Na}
\end{align}
and
\begin{align}
&F_{T/L}^{(b)} \, \vec{\epsilon}_{\lambda,\text{T/L}}\cdot\vec{S}_{M_J^\prime
  M_J} =
3 \, \langle M_J' |\, (\vec{\tau}_1\cdot\vec{\tau}_2-\tau_1^z\tau_2^z)
\nonumber\\
&\times \frac{\big[(\vec{p}_{12}-\vec{p}_{12}^{\;\prime}-
\vec{k}_\gamma/2)\cdot (\vec \sigma_1 + \vec \sigma_2 ) \big] }
{\big[(\vec{p}_{12}-\vec{p}_{12}^{\;\prime}- \vec{k}_\gamma/2 )^2+M_\pi^2 \big]
\big[\vec{p}_{12}-\vec{p}_{12}^{\;\prime}+\vec{k}_\gamma/2 \big]^2} \nonumber \\
&\times \big[
\vec{\epsilon}_{\lambda,\text{T/L}}
\cdot (\vec{p}_{12}-\vec{p}_{12}^{\;\prime}) \big]\, 
|M_J\rangle_\psi~,  
\label{eq:valF2Nb}
\end{align}
where $\vec{p}_{12} =(\vec{k}_1-\vec{k}_2)/2$ and
$\vec{p}_{12}^\prime=(\vec{k}_1^\prime-\vec{k}_2^\prime)/2$ 
are the initial and final Jacobi momenta
of nucleons 1 and 2, respectively. The integral for the 
form factors $F_{T/L}^{(a)}$ contains an integrable singularity which can
be removed by an appropriate variable transformation. Then, 
the form factors can 
be evaluated using Monte Carlo integration in the same way as the 
form factors for the single-nucleon contribution.
Our results for $F_{T/L}^{(a)}-F_{T/L}^{(b)}$ are given
in Table~\ref{tab:Fab}.
\renewcommand{\arraystretch}{1.3}
\begin{table}[t]
\begin{center}
\begin{tabular}{|c||c|c|}
\hline 
nucleus & $^3$He & $^3$H \\
\hline
$F_{T}^{(a)} -F_{T}^{(b)}$ [fm$^{-1}$]
&  $-29.3(2)(1)$ & $-29.7(2)(1)$ \\
$F_{L}^{(a)} -F_{L}^{(b)}$ [fm$^{-1}$]
&  $-22.9(2)(1)$ & $-23.2(1)(1)$ \\
\hline
\end{tabular}
 \caption{Numerical results for the form factors 
  $F_{T/L}^{(a)} -F_{T/L}^{(b)}$ parametrizing two-body contributions in units of fm$^{-1}$.
   The first error is an estimate of the theory error
   from higher orders in chiral EFT while the second error
   is the statistical error from the Monte Carlo integration.}
  \label{tab:Fab}
\end{center}
\end{table}
The first error is again the theory error estimated from the cutoff variation
in the chiral interaction as described above.
The second error is the statistical error from the Monte Carlo integration
which is about half the size of the theory error.

\section{Results and Discussion}

We are now in the position to evaluate the
nuclear S-wave multipoles. They are given as the
sum of the one- and the two-nucleon contributions given
in Eqs.~(\ref{eq:mult1N}, \ref{eq:mult2N}) in the previous section,
\beqa
E_{0+} &=& E_{0+}^{\rm 1N}+E_{0+}^{\rm 2N}~,\nonumber\\
L_{0+} &=& L_{0+}^{\rm 1N}+L_{0+}^{\rm 2N}~.
\eeqa
Using the values for the one- and two-body form factors 
in Tables \ref{tab:FTL} and \ref{tab:Fab} together with
the subleading chiral perturbation theory results for the 
single-nucleon multipoles at ${\cal O}(p^4)$
\cite{Bernard:1994gm,Bernard:2001gz}
\beqa
E_{0+}^{\pi^0 p} &=& -1.16 \times 10^{-3}/M_{\pi^+}~,\nonumber\\
E_{0+}^{\pi^0 n} &=& +2.13 \times 10^{-3}/M_{\pi^+}~,\nonumber\\
L_{0+}^{\pi^0 p} &=& -1.35 \times 10^{-3}/M_{\pi^+}~,\nonumber\\
L_{0+}^{\pi^0 n} &=& -2.41 \times 10^{-3}/M_{\pi^+}~,
\label{eq:1Nmultipoles}
\eeqa
we obtain for the threshold multipoles on $^3$He and on $^3$H
the values in Table~\ref{tab:multip}.
\renewcommand{\arraystretch}{1.3}
\begin{table*}[t]
\begin{center}
\begin{tabular}{|c|c||c|c|c|}
\hline 
nucleus & multipole & 1N & 2N & total \\
\hline
$^3$He & $E_{0+}$  [$10^{-3}/M_{\pi^+}$]
& +1.71(4)(9) & -3.95(3) & -2.24(11)\\  
& $L_{0+}$ [$10^{-3}/M_{\pi^+}$]
& -1.89(4)(9) & -3.09(2) & -4.98(12)\\
\hline  
$^3$H & $E_{0+}$ [$10^{-3}/M_{\pi^+}$] 
& -0.93(3)(5) & -4.01(3) & -4.94(7)\\  
& $L_{0+}$ [$10^{-3}/M_{\pi^+}$]
& -0.99(4)(5) & -3.13(1) & -4.12(7)\\
\hline
\end{tabular}
 \caption{Numerical results for the threshold multipoles $E_{0+}$
and $L_{0+}$ on $^3$He and $^3$H. The error estimates are explained in the
text.}
  \label{tab:multip}
\end{center}
\end{table*}
For the 1N contribution, the first error is the theory error
from higher orders in chiral EFT estimated from the cutoff variation
as explained above,
while the second error is from the 5\% uncertainty of the one-nucleon
amplitudes. In the case of the 2N contribution, only the 
theory error is given. The total error is obtained by adding the theory
error and the uncertainty of the one-nucleon amplitudes in quadrature.
As noted before, the total error is
dominated by the uncertainty in the single nucleon amplitudes.
We stress that our estimate of the theory error is only a lower bound. 

One observes that the multipoles get a large contribution from the 
two-body terms.
This behavior is similar to the deuteron case \cite{Beane:1997iv}.
For example, in the case of $E_{0+}$ for $^3$He, the proton contribution is
$-0.01$, 
the neutron one is 1.72 while the two-body contribution is $-3.95$
in the canonical units of $ 10^{-3}/M_{\pi^+}$.
There is, however, still a large sensitivity to the 
single-neutron contribution.

\begin{figure}[ht]
        \begin{center}
        \includegraphics*[width=0.95\linewidth]{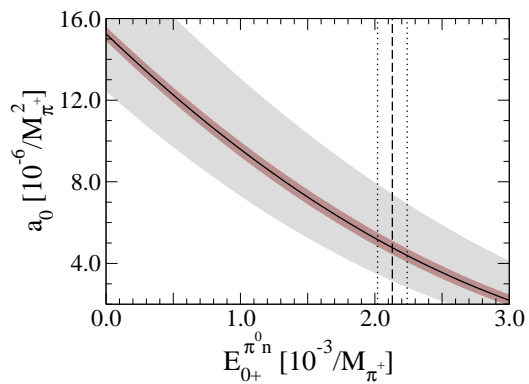}
        \end{center}
        \vspace{-0.2cm}
        \caption{Sensitivity of $a_0$ for $^3$He in units of 
        $ 10^{-6}/M^2_{\pi^+}$
        to the single-neutron multipole $E_{0+}^{\pi^0 n}$
        in units of $ 10^{-3}/M_{\pi^+}$. The vertical dashed line
        gives the CHPT prediction for $E_{0+}^{\pi^0 n}$ and the 
        vertical dotted lines indicate the 5\% error in the prediction.
        The inner shaded band indicates the theory error estimated
        from the cutoff variation as described in the text. 
        The outer shaded band corresponds to
        a 10\% uncertainty in the 2N contribution.}
        \label{fig:E0psq}
        \end{figure}

The corresponding threshold S-wave cross section 
for pion photoproduction $a_0$ is given by
\beq
a_0 =\left.\frac{|\vec{k}_\gamma|}{|\vec{q}_\pi|}
\frac{d\sigma}{d\Omega}\right|_{\vec{q}_\pi=0}=\left|E_{0+}\right|^2~.
\eeq
The longitudinal multipole $L_{0+}$ contributes only in
electro-production. 
The corresponding threshold cross
section contains an extra term $\sim \left|L_{0+}\right|^2$.
From here on, we will, however, concentrate on photoproduction. 
In Fig.~\ref{fig:E0psq}, we illustrate the sensitivity of 
$a_0$ to the single-neutron multipole $E_{0+}^{\pi^0 n}$.
The inner shaded band indicates the theory error estimated
from the cutoff variation as described above. The outer shaded band illustrates the
effect of a 10\% uncertainty in the 2N contribution. This corresponds to the 
size of the correction to the 2N contribution at the next order in the 
deuteron case~\cite{Beane:1997iv}.
The vertical dashed line indicates the CHPT prediction 
$E_{0+}^{\pi^0 n}= 2.13 \times 10^{-3}/M_{\pi^+}$.
Changing this value by $\pm20\%$ leads to changes in $a_0$ 
of about $\pm30\%$.
Thus, the $^3$He nucleus appears to be a very promising target to test
the CHPT prediction for $E_{0+}^{\pi^0 n}$. 
On the contrary, neutral pion production on $^3$H is rather 
insensitive to $E_{0+}^{\pi^0 n}$:
a variation of $E_{0+}^{\pi^0 n}$ from 0 to 3 changes  $a_0$
only by $1\%$.

Next we compare our predictions with the available data. 
The consistency of the CHPT prediction for the single-neutron 
multipole with the measured S-wave threshold amplitude on the
deuteron from Saclay and Saskatoon
is well established, see Refs.~\cite{Beane:1997iv,Bernard:2007zu}.
The reanalyzed measurement of the S-wave amplitude for $^3$He
at Saclay gives
$E_3 = (-3.5\pm 0.3)\times 10^{-3}/M_{\pi^+}$ \cite{Argan:1980zz,Argan:1987dm},
which is related to $a_0$ according to 
\beq
\left|E_{0+}\right|^2 = |E_3|^2 \left|\frac{F_T^{S-V}}{2} \right|^2 \left(
\frac{1+M_\pi/m_N}{1+M_\pi/3m_N}\right)^2~.
\eeq
Here, we have approximated the $A=3$ body form factor ${\cal F}_A$
of Argan et al.~\cite{Argan:1987dm} by the numerically dominant form factor
$F_{T}^{S-V}$ 
for $^3$He, cf.~Tab.~\ref{tab:FTL}. This results in 
\beq
E_{0+} = (-2.8 \pm 0.2) \times 10^{-3}/M_{\pi^+}~,
\eeq
assuming the same sign as for our $^3$He prediction in
Table~\ref{tab:multip}.
In magnitude,
the extracted value is about 25\% above 
the predicted one. Given the
model-de\-pen\-dence that is inherent to the analysis of Ref.~\cite{Argan:1987dm},
it is obvious that a more precise measurement using  CW beams and modern
detectors is very much called for.

\section{Summary and Outlook}

In this letter, we have presented a calculation of neutral pion 
production off $^3$H and  $^3$He at threshold to leading one-loop order
for the production operator in the framework of chiral nuclear effective field theory. 
We used the chiral wave functions of
Refs.~\cite{Epelbaum:2003gr,Epelbaum:2003xx} 
which are consistent with the 
pion production operator to calculate the S-wave 3N multipoles $E_{0+}$ and $L_{0+}$.
To this order, the production operator gets both
one- and two-body contributions. Our calculation shows that the two-body
contributions are of the same order of magnitude as the one-body
contributions. A similar behavior was observed in  
the deuteron case~\cite{Beane:1997iv}.

The theoretical uncertainty resulting from the cutoff variation in the
employed wave functions appears to be small (of the order of 
$3\%$).  The dominant theoretical error at this order stems
from the threshold pion production amplitude off the proton and the
neutron, which is estimated to be about 5\%.

We explored the sensitivity 
of neutral pion photoproduction on $^3$He to the elementary neutron multipole
$E_{0+}^{\pi^0 n}$ and found a large sensitivity. This makes $^3$He
a promising target to test the counterintuitive CHPT prediction
for  $E_{0+}^{\pi^0 n}$~\cite{Bernard:1994gm,Bernard:2001gz}.
The cutoff variation estimate leads to a very small error for the 2N 
contribution. If the error of this contribution is artificially
enlarged by a factor of 10, the extraction of the neutron multipole
is still feasible experimentally.

We have shown that our prediction for the $^3$He S-wave multipole
$E_{0+}$ is roughly consistent with the  value deduced from the old
Saclay measurement of the threshold cross section \cite{Argan:1987dm}.
A new measurement using modern technology and better
methods to deal with  few-body dynamics is urgent\-ly called for. 

There are many natural extensions of this work. They include investigating
higher orders, pion production above threshold, the extension to virtual photons
and pion electroproduction, production of charged pions, and 
considering heavier
nuclear targets such as $^4$He. 
Further work in these directions is in progress.

\section*{Acknowledgements}
We thank Andreas Nogga for providing us with the chiral 3N 
wave functions and Hermann Krebs for discussions.
Financial support by the Deutsche Forschungsgemeinschaft 
(SFB/TR 16, ``Subnuclear 
Structure of Matter''),
by the European Community Research Infrastructure Integrating Activity 
``Study of Strongly Interacting Matter''
(acronym HadronPhysics2, Grant A\-gree\-ment n.~227431) under the 7th
Framework Programme of the EU and by 
the European Research Council (acronym NuclearEFT, ERC-2010-StG 259218)
is gratefully
acknowled\-ged.

\end{document}